\documentclass{ws-procs9x6}
\usepackage{bbm}

\def\rz{{\mathbbm{R}}}

\begin{document}

\title{Breathers in a model of a polymer with secondary structure}

\author{MICHAEL KASTNER and J.~TOM\'AS L\'AZARO%
\footnote{Permanent address: Departament de Matem\'atica Aplicada I, Universitat Polit\'ecnica de Catalunya, Diagonal 647, E-08028 Barcelona, Spain.}
}

\address{I.N.F.M., UdR di Firenze, \\
Via G.~Sansone 1, \\ 
50019 Sesto Fiorentino, Italy\\ 
E-mail: kastner@fi.infn.it, jose.tomas.lazaro@upc.es}

\maketitle

\abstracts{
A simple model of a polymer is considered: a chain of (different) point masses, connected by harmonic springs, embedded in two dimensional space. In order to determine conditions for existence and stability of breather excitations, the method of numerical continuation of a breather solution from the anticontinuous limit is employed. Approaching the limit of equal masses, stable breather solutions are found only within an extremely narrow band of frequencies.
}

\section{Introduction}
The search for discrete breather solutions in simple models of polymers with a secondary structure might be regarded as a step towards an understanding of the role that local excitations might play for the functionality of proteins. A minimal model should include the following two features:
\begin{enumerate}
\item a one-dimensional system (chain) is embedded in $d$-dimensional space ($d>1$),
\item inter-particle interactions involve at least $d$ neighbours in order to obtain a secondary structure.
\end{enumerate}

\section{The model}
We use a slightly modified version of a model proposed by Zolotaryuk {\em et al.}:\cite{ZoChriSa} a chain of $N$ classical particles (point masses) $m_i$, $i=1,...,N$, which interact by means of linear forces between nearest and next-nearest neighbours. The Hamiltonian function of the system is
\begin{eqnarray*}
{\mathcal H}&:&\rz^{4N}\to\rz\\
&&({\bf q_1},...,{\bf q_N},{\bf p_1},...,{\bf p_N})\mapsto\sum_{i=1}^N \frac{{\bf p_i}^2}{2m_i}+\\
&&\quad+\sum_{i=1}^{N-1}U(|{\bf q_i}-{\bf q_{i+1}}|)+\\
&&\quad+\sum_{i=1}^{N-2}V(|{\bf q_i}-{\bf q_{i+2}}|).
\end{eqnarray*}
%\begin{figure}[th]
%\centerline{\epsfxsize=4.0in\epsfbox{chain.eps}}   
%\caption{Chain of particles with nearest and next-nearest neighbour interactions.}
%\end{figure}
The particles are allowed to move in the Euklidian plane, i.e., ${\bf q_i}, {\bf p_i}\in\rz^2$. For closest similarity to real polymers, free boundary conditions are employed. In contrast to the original model of Zolotaryuk {\em et al.}, we allow different values for the masses $m_i$. This facilitates the construction of an 'anticontinuous limit' as explained (and required) in Sec.\ 3.1.

For simplicity, harmonic interaction potentials $U(x)=V(x)=\frac{1}{2}x^2$ are chosen between nearest as well as next-nearest neighbouring particles. Despite the linearity (in the inter-particle distance) of the forces, the geometry of the chain gives rise to effective nonlinearities in the equations of motion, and therefore existence of breather solutions is not ruled out from the outset.

The system displays a secondary structure, for example a zig-zag chain (or 2-helix) as illustrated below. (Note, however, that the 'equilibrium configuration' of the system is not unique!)
\begin{figure}[th]
\centerline{\epsfxsize=3.5in\epsfbox{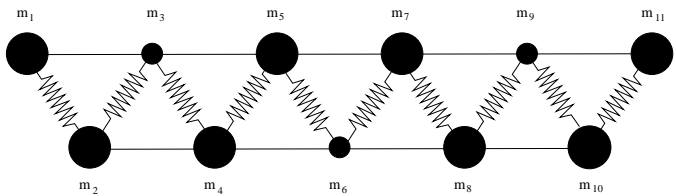}}
\caption{Zig-zag chain.\label{zigzag}}
\end{figure}

\section{Existence of breathers and stability conditions}
We are interested in the following question: {\em In which way do the parameters of the system (like oscillation frequencies or particle masses) have to be tuned in order to find breather solutions?}

\subsection{Numerical calculation from the anticontinuous limit}
A numerical method due to Mar\'{\i}n and Aubry\cite{MaAub} is used in order to determine the existence region of breather solutions: A (known) breather solution of the system for certain parameter values is continued to other (not too different) parameter values by means of Newton's method. Such a continuation is possible, in principle also analytically, for breather oscillation frequencies which are non-resonant with the spectrum of the linearized equations of motion of the system. This allows us to obtain breather solutions for, e.g., arbitrary mass ratios and frequencies, provided that
\begin{enumerate}
\item such solutions exist, and
\item a breather solution is known from which continuation up to the desired parameter values is feasible.
\end{enumerate}
As a starting point for such a continuation, a certain limiting value (the so-called anticontinuous limit) of a parameter is taken, for which existence of breather solutions is known. Here, a diatomic chain is considered, where two particles of mass $M\geq1$ are followed by one particle of mass $m=1$:
\[
MMmMMmMMmMMm...
\]
Existence of localized oscillations (breathers) is obvious in the limit of zero mass ratio, $\frac{m}{M}\to0$, for example: all particles are at rest but a single one of mass $m$, which is coupled only to particles of mass $M\to\infty$. This solution can be used as a starting point for continuation.

\subsection{An ``existence and stability diagram''}

\begin{figure}[tbh]
\vspace{-0in}
\centerline{\epsfxsize=4.5in\epsfbox{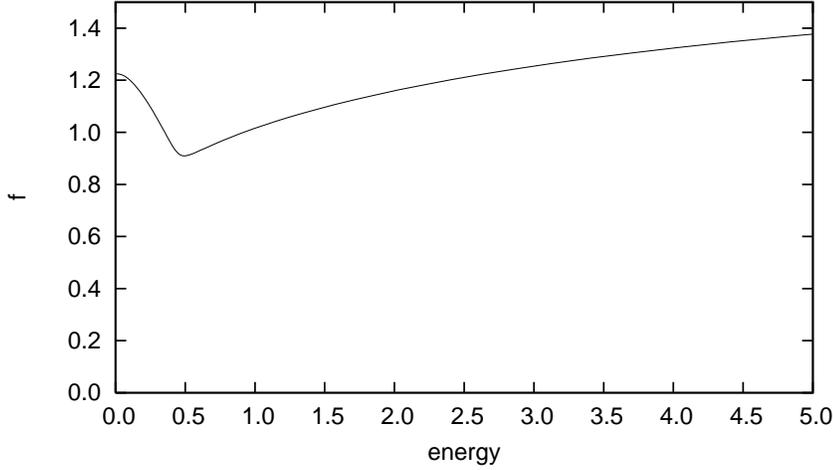}}   
\caption{Frequency vs.\ energy of a single particle of mass $m=1$ in the anticontinuous limit.\label{freq_en}}
\end{figure}

\begin{figure}[b]
\vspace{-0in}
\centerline{\epsfxsize=5in\epsfbox{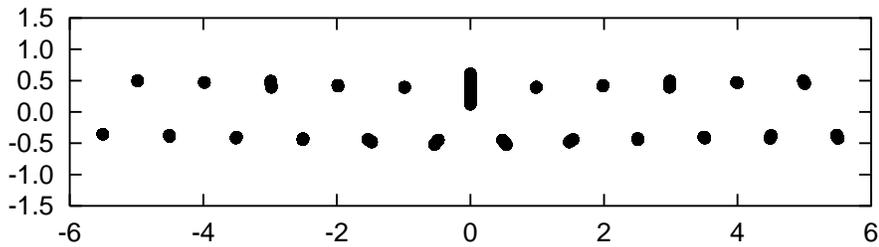}}   
\caption{Orbits of the particles of a symmetric breather solution located at the central position for mass ratio $\frac{m}{M}=\frac{2}{7}$ and frequency $f=0.212$.\label{DBplot}}
\end{figure}

The number of parameters present in the system is still quite large. As an example, existence and stability regions of discrete breathers are investigated for a particular choice of parameters:
\begin{itemize}
\item A chain of 23 particles.
\item The set of initial conditions in the anticontinuous limit is restricted: all but the centre particle are at rest. The centre particle moves on the reflection symmetry line of the zig-zag chain. (Consider a vertical oscillation of particle 6 in Figure \ref{zigzag}, while all the other particles are at rest.)
\item For an oscillating particle in the anticontinuous limit, the frequency is a non-monotonic function of the energy (see Fig.\ \ref{freq_en}). We restrict the initial conditions to energies corresponding to the decreasing (left hand) part of $f(energy)$.
\end{itemize}
In order to visualize such a breather solution, the orbits of the particles of a breather solution are plotted in Fig.\ \ref{DBplot} for mass ratio $\frac{m}{M}=\frac{2}{7}$ and frequency $f=0.212$.

\begin{figure}[tbh]
\centerline{\epsfxsize=4.0in\epsfbox{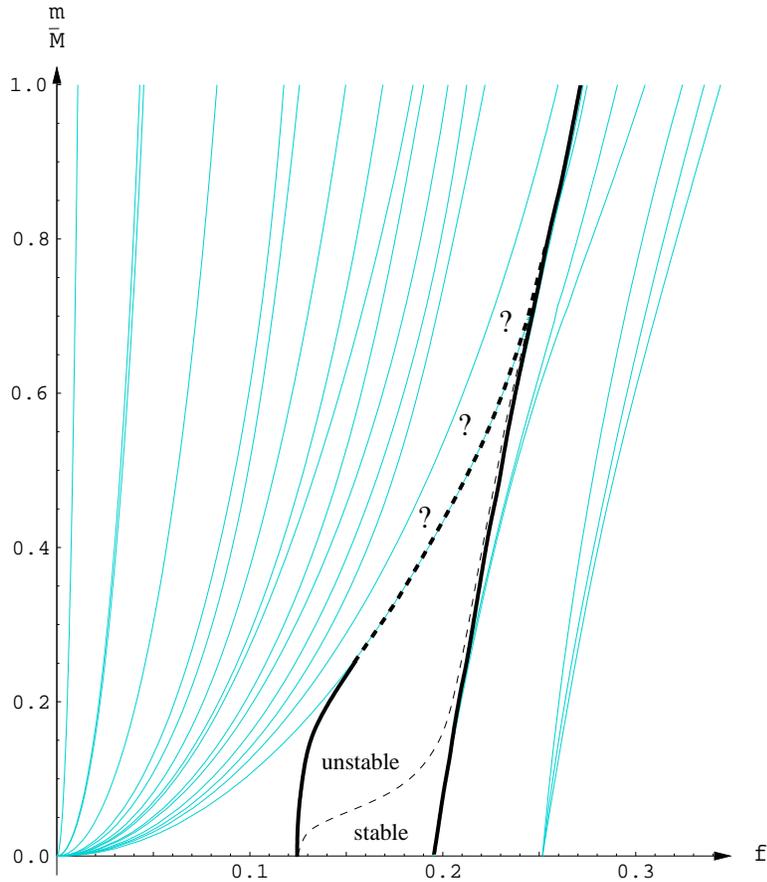}}   
\caption{Existence and stability regions of symmetric breather solutions in the parameter space of breather frequencies $f$ and mass ratios $\frac{m}{M}$. The bottom line corresponds to the anticontinuous limit. The thin gray lines are the eigenfrequencies of the linearized equations of motion. Under the conditions specified above, breather solutions are found for parameter values lying between the solid black lines.\newline
The solid black line on the right hand side coincides with an eigenfrequency of the linearized equations of motion. Approaching this boundary, the breather amplitude goes to zero. On the left hand side, the region of breather existence is more difficult to determine, mainly because of the following reason: the concept of localization is somewhat ill-defined in finite systems. Numerically, breather solutions can be continued across the bold dashed line coinciding, again, with an eigenfrequency of the linearized equations of motion. The localization of the solutions found in this region, however, is so weak, that a distinction between extended and localized modes does not seem feasible.\newline
In between these two boundaries breather solutions exist, although, for larger mass ratios $\frac{m}{M}$, stable breathers appear only within an extremely narrow frequency range.\label{ex+stab}}
\end{figure}

Exemplarily for the above choices of particle number and initial conditions, Figure \ref{ex+stab} shows the regions in the parameter space of frequencies and mass ratios, for which symmetric breather solutions can be found.

\subsection{Stability of breathers}
In the anticontinuous limit, all breather solutions are stable. For non-zero mass ratios $\frac{m}{M}$, there are regions of stable and unstable breathers. Note that both, stable and unstable breathers, are obtained by continuation from the same solution in the anticontinuous limit, and continuation across the stability boundary does not cause any numerical difficulties. The regions of (linearly) stable and unstable breathers are indicated in Figure \ref{ex+stab}.

In contrast to the original work [2], the numerical continuation was performed with respect to {\em two parameters}, namely the mass ratio and the frequency. From Figure \ref{ex+stab} it becomes clear, why this process is essential: for masses $M\lesssim 10$, stable breather solutions cannot be found by continuing a solution from the anticontinuous limit while keeping the frequency fixed.

\subsection{Other parameter values}
Varying some parameters of the model under investigation, quantitative changes occur, while no qualitative modifications of the phase diagram have been observed:
\begin{itemize}
\item Due to the localized character of the oscillation, increasing the number of particles does not lead to significant changes (maybe apart from the case of mass ratio $\frac{m}{M}$ close to unity, where localization is less distinct).
\item Enlarging the coupling constant of the nearest neighbour interaction (while keeping the next-nearest neighbour coupling constant to unity), the lower bound on the accessible frequencies can be shifted towards zero.
\item Breather existence and stability can be strongly influenced by the shape of the interaction potentials $U$ and $V$ in the Hamiltonian. In particular the softening (as in our case) or hardening property of the potentials often gives rise to entirely different behaviour of the respective systems.
\end{itemize}

\section*{Acknowledgments}
We would like to thank Roberto Livi for arousing our interest in the topic of discrete breathers and for stimulating discussions. The work was supported by EU contract HPRN-CT-1999-00163 (LOCNET network).

\end{document}